\documentclass[aps,prl,twocolumn,superscriptaddress]{revtex4}
\usepackage{graphicx}
\usepackage{url}
\usepackage{natbib}
\usepackage{amsmath,amssymb}

\begin{document}

\title{Crystal structure of the pyrochlore oxide superconductor KOs$_2$O$_6$}
\author{Jun-Ichi Yamaura, Shigeki Yonezawa, Yuji Muraoka, Zenji Hiroi}
\affiliation{Institute for Solid State Physics, The University of Tokyo, Kashiwa, Chiba 277-8581, Japan}
\date{11 November 2005}
\begin{abstract}
We report the single-crystal X-ray analysis of the structure of the pyrochlore oxide superconductor KOs$_2$O$_6$. The structure was identified as the $\beta$-pyrochlore structure with space group $Fd\bar{3}m$ and lattice constant $a$ = 10.089(2)~\AA\ at 300~K: the K atom is located at the 8$b$ site, not at the 16$d$ site as in conventional pyrochlore oxides. We found an anomalously large atomic displacement parameter $U_\mathrm{iso}$ = 0.0735(8)~\AA$^2$ at 300~K for the K cation, which suggests that the K cation weakly bound to an oversized Os$_{12}$O$_{18}$ cage exhibits intensive rattling, as recently observed for clathrate compounds. The rattling of A cations is a common feature in the series of $\beta$-pyrochlore oxide superconductors AOs$_2$O$_6$ (A = Cs, Rb and K), and is greatest for the smallest K cation.
\end{abstract}

\maketitle

Pyrochlore oxides exhibit many fascinating phenomena, such as metal--insulator transition, superconductivity, magnetic frustration, colossal magnetoresistance, and ferroelectrics~\cite{Subramanian}. The first pyrochlore oxide superconductor Cd$_2$Re$_2$O$_7$ with $T_\mathrm{c}$ = 1.0~K was found in 2001 and is now believed to be a conventional BCS-type superconductor~\cite{Hanawa}. Recently, a new family of pyrochlore oxide superconductors AOs$_2$O$_6$ was discovered, with $T_\mathrm{c}$ = 3.3, 6.3 and 9.6~K for A = Cs, Rb and K, respectively~\cite{Yonezawa1,Yonezawa2,Yonezawa3,Bruhwiler}. Several experimental results for KOs$_2$O$_6$ with the highest $T_\mathrm{c}$ have indicated unconventional features of superconductivity; the absence of a coherence peak in NMR experiments~\cite{Arai}, a large upper critical field $H_\mathrm{c2}\sim 38$~T~\cite{Hiroi4}, and an anisotropic order parameter from muon spin rotation ($\mu$SR) experiments~\cite{Koda}. In contrast, experimental results for A = Cs and Rb have indicated rather conventional features, although the mechanism of superconductivity is likely the same among the three compounds~\cite{Bruhwiler,Arai}.

In ordinary pyrochlore oxides with the chemical formula A$_2$B$_2$O$_6$O', the A cation, which is a large, rare-earth alkali metal, alkaline earth metal or post-transition metal cation, lies in a distorted (6O+2O') coordination, while the B cation, which is a small transition metal cation, forms a slightly distorted BO$_6$ octahedron. In the space group $Fd\bar{3}m$ with the origin chosen at the B site, the A, B, O and O' atoms occupy the 16$d$, 16$c$, 48$f$ and 8$b$ Wyckoff positions, respectively. The crystal structure of AOs$_2$O$_6$ was preliminarily reported to be the $\beta$-pyrochlore structure, in which the alkali metal cation A moves from the 16$d$ site to the 8$b$ site, replacing the O' atom of the conventional ($\alpha$-) pyrochlore oxides, while the B(Os)--O network remains intact~\cite{Yonezawa1,Yonezawa2,Yonezawa3,Bruhwiler}. A similar structure is observed in so-called defect pyrochlore oxides. For instance, in ANbWO$_6$ with A = Cs, Rb and K, large Cs and Rb atoms occupy the 8$b$ site, while the smaller K atom lies at the 32$e$ site away from the 8$b$ site~\cite{Barnes}. Moreover, it is known that in KNbWO$_6$ a water molecule can be intercalated near the 8$b$ site, pushing a K atom towards the 16$d$ site. Structural refinements have been carried out for A = Cs and Rb by Rietveld analysis using synchrotron X-ray powder diffraction~\cite{Yonezawa4}. The lattice constants at room temperature are 10.15250(4) and 10.11760(4)~\AA, and the atomic coordinates $x$ of the 48$f$ oxygen are 0.3146(3) and 0.3180(3) for CsOs$_2$O$_6$ and RbOs$_2$O$_6$, respectively. Remarkably large atomic displacement parameters are observed for alkali metal atoms: $U_\mathrm{iso}$ = 0.025~\AA$^2$ and $U_\mathrm{iso}$ = 0.043~\AA$^2$ for Cs and Rb, respectively, which are much greater than $U_\mathrm{iso}$~$\sim$~0.009~\AA$^2$ for Os atoms. On the other hand, detailed structural analysis for KOs$_2$O$_6$ has not yet been reported because of difficulty in obtaining a single-phase polycrystalline sample and because of the hygroscopic nature of the powdered sample. In this paper, we report the first X-ray structural analysis of KOs$_2$O$_6$ using a fresh single crystal.

A single crystal was synthesized from KOsO$_4$ and Os in a sealed quartz tube with an appropriate amount of AgO for oxygen supply at 450~${}^\circ\mathrm{C}$ for 48~h. Energy-dispersive X-ray analysis in a scanning electron microscope indicated that the atomic ratio of K to Os is approximately equal to 0.5. Magnetic susceptibility was measured in a SQUID magnetometer (Quantum Design MPMS). Structural analyses were performed at 300, 200 and 100~K using a fresh crystal of 0.118$\times$0.070$\times$0.052~mm in a three-circle diffractometer equipped with a CCD area detector (Bruker SMART APEX) and a nitrogen-flow cooler. Oscillation photographs were taken at lower temperatures from 100 to 5~K using a curved imaging plate (MacScience DIP 320V) and a helium closed-cycle refrigerator. Monochromatic Mo-$K\alpha$ radiation was used as the X-ray source. Data integration and global-cell refinements were performed with SAINT using data in the range 2$\theta$ = 4$-$107$^\circ$, and absorption correction was applied using SADABS~\cite{SADABS}. Symmetry equivalent reflections were averaged to produce unique reflections ($R_\mathrm{int}$ = 0.051$-$0.055). Structural parameters were refined by the full-matrix least-squares method using the TEXSAN program.

\begin{figure}[htb]
\centering
\includegraphics[height=5.6cm,width=7.4cm]{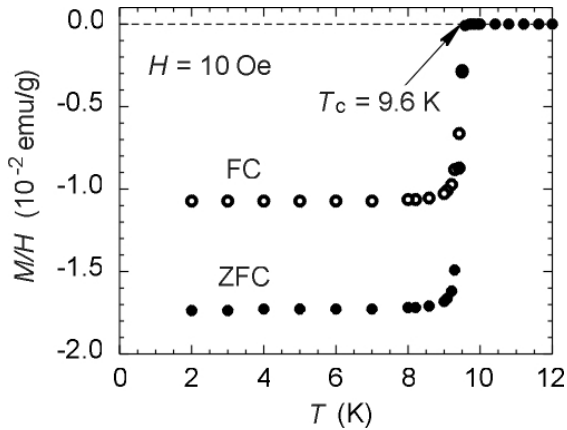}
     \caption{Temperature dependence of magnetic susceptibility measured on a single crystal of KOs$_2$O$_6$ in a magnetic field of 10~Oe. ZFC ($\bullet$) and FC ($\circ$) indicate the zero-field cooling and the field cooling experiments, respectively.}
     \label{fig:5}
\end{figure}

Figure~\ref{fig:5} shows the temperature dependence of magnetic susceptibility measured in a magnetic field of 10~Oe using another single crystal obtained from the same batch as used in the X-ray experiments. A superconducting transition was observed below $T_\mathrm{c}$ = 9.6~K, as previously reported~\cite{Yonezawa1}. The volume fraction estimated at 2~K in the zero-field cooling experiment is 144~\%, which means that nearly 100~\% volume of the crystal becomes superconductivity, taking into account the demagnetization correction.

\begin{figure}[htb]
\centering
\includegraphics[height=5cm,width=7.6cm]{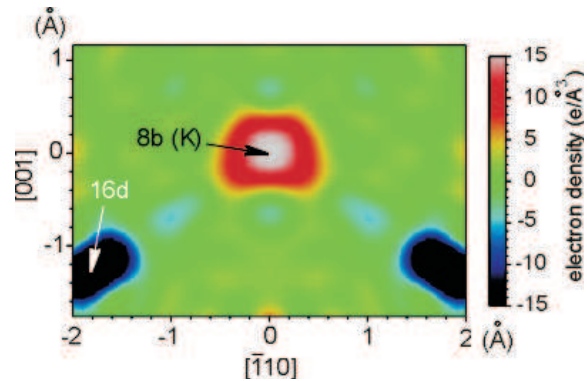}
      \caption{Difference Fourier map at 300~K on the (1 1 0) plane around the 8$b$ site (3/8,~3/8,~3/8) with the Os and O atoms subtracted. The 16$d$ site (1/2,~1/2,~1/2) lies at the midpoint between neighboring 8$b$ sites, and the 32$e$ site ($x$,~$x$,~$x$) at the midway between the 8$b$ and 16$d$ sites.}
      \label{fig:1}
\end{figure}

The unit cell determined at 300, 200 and 100~K was $a$ = 10.089(2), 10.086(2) and 10.083(2)~\AA, respectively. The systematic absence observed is consistent with space group $Fd\bar{3}m$. No additional reflections were observed over the temperature range 5$-$300~K, suggesting the absence of structural transitions, in contrast to the case of related $\alpha$-pyrochlore oxide Cd$_2$Re$_2$O$_7$, for which two successive structural transitions of decreasing symmetry occur~\cite{Yamaura}. In the first step to determine the crystal structure at 300~K, the positions of the Os and O atoms were fixed at the 16$c$ and 48$f$ sites, respectively, according to the direct method~\cite{SIR}. Then, we carefully tried to fix the position of the K atom using the Fourier method. Figure~\ref{fig:1} shows a difference Fourier map on the (1 1 0) plane around the 8$b$ site, which was obtained by subtracting the structure factor calculated for the Os and O atoms from that observed at 300~K. Apparently, the K atom is found at the 8$b$ (3/8,~3/8,~3/8) position, and both the 32$e$ ($x$,~$x$,~$x$) and 16$d$ (1/2,~1/2,~1/2) positions are empty. The occupancy of the K atom is refined as 1.03(12), suggesting that there are no vacancies within the experimental error. Moreover, no evidence of water insertion such as reported for KNbWO$_6$ was detected in the present compound~\cite{Barnes}. Essentially the same results were obtained at 200 and 100~K. The final $R$[$F$] ($R_\mathrm{w}$[$F^2$]) values were 0.028 (0.031), 0.027 (0.033) and 0.024 (0.029) for $I>3.0\sigma$($I$) at 300, 200 and 100~K, respectively. Therefore, the crystal structure of KOs$_2$O$_6$ was determined to be the ideal $\beta$-pyrochlore structure. The refined structure parameters are given in Table~\ref{tbl:1}.

\begin{table}[h]
\caption{Atom coordinates and isotropic atomic displacement parameters for KOs$_2$O$_6$ at 300, 200 and 100~K}
\label{tbl:1}
\begin{center}
\begin{tabular}{cccccc}
\hline
Atom  & Position & $x$       & $y$   & $z$   & 100$U_\mathrm{iso}$ (\AA$^2$) \\
\hline
300.0(1)~K & \multicolumn{3}{l}{~~$a$ = 10.089(2) \AA} \\
K     &  8$b$ & 0.375     & 0.375 & 0.375 & 7.35(8)  \\
Os    & 16$c$ & 0         & 0     & 0     & 0.435(2) \\
O     & 48$f$ & 0.3145(8) & 0.125 & 0.125 & 1.24(8)  \\
\\
200.0(1)~K & \multicolumn{3}{l}{~~$a$ = 10.086(2) \AA} \\
K     &  8$b$ & 0.375     & 0.375 & 0.375 & 5.46(6)  \\
Os    & 16$c$ & 0         & 0     & 0     & 0.353(2) \\
O     & 48$f$ & 0.3140(7) & 0.125 & 0.125 & 1.23(7)  \\
\\
100.0(1)~K & \multicolumn{3}{l}{~~$a$ = 10.083(2) \AA} \\
K     &  8$b$ & 0.375     & 0.375 & 0.375 & 3.89(4)  \\
Os    & 16$c$ & 0         & 0     & 0     & 0.286(2) \\
O     & 48$f$ & 0.3155(6) & 0.125 & 0.125 & 0.90(5)  \\
\hline
\end{tabular}
\end{center}
\end{table}

\begin{figure}[htb]
\centering
\includegraphics[height=11.7cm,width=7.7cm]{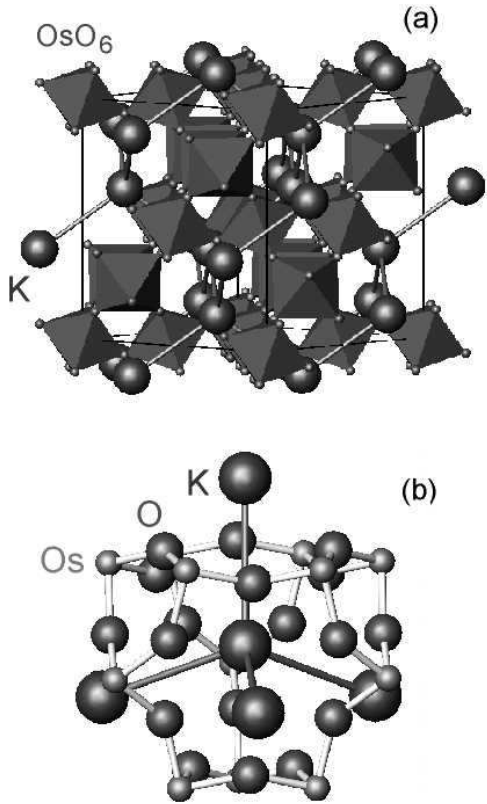}
     \caption{Crystal structure of $\beta$-pyrochlore oxide KOs$_2$O$_6$. Corner-shared OsO$_6$ octahedra form a three-dimensional network, and the K atoms at the 8$b$ site are located in the center of the Os$_{12}$O$_{18}$ cage as shown in (b). The K atoms hypothetically form a diamond sublattice with long K--K bonds through the four channels of the Os$_{12}$O$_{18}$ cage along the $\langle 111 \rangle$ direction.}
     \label{fig:2}
\end{figure}

Figure~\ref{fig:2} illustrates the crystal structure of KOs$_2$O$_6$. There are eight formula units per unit cell. The OsO$_6$ octahedra share their corners to form a three-dimensional network. The atom coordinate $x$ of O (48$f$) is a unique parameter that determines the distortion of an OsO$_6$ octahedron. The deviation from the special value of $x$ = 0.3125 for the ideal octahedron leads to a trigonal distortion with an equal Os--O distance. The values of $x$ = 0.3140$-$0.3155 observed for KOs$_2$O$_6$ are much smaller than those for other pyrochlore oxides with $x$ = 0.32$-$0.33~\cite{Subramanian}. This means that there is a small distortion of the octahedron, with internal O--Os--O angles of 90.6$-$91.2${}^\circ$ and 88.8$-$89.4${}^\circ$ and a large Os--O--Os angle of 139.4$-$140.2${}^\circ$ between neighboring octahedra, as listed in Table~\ref{tbl:2}. The octahedral distortion seems to increase slightly at low temperature. The valence of Os is estimated to be +5.5 from bond valence sum calculations using a bond valence parameter of 1.868 for Os$^{5+}$ in the empirical formula and the Os--O bond distances obtained at 300~K~\cite{Yonezawa4,Brese}. This is in good agreement with the formal valence of +5.5 expected from the purely ionic model.

\begin{table}[h]
\caption{Os--O distance, O--Os--O bond angles in the octahedron and Os--O--Os bond angles at 300, 200 and 100~K}
\label{tbl:2}
\begin{center}
\begin{tabular}{cccc}
\hline
$T$ (K)  & Os--O (\AA) & O--Os--O (${}^\circ$) & Os--O--Os (${}^\circ$) \\
\hline
300 & 1.898(3)   & 90.8(3)/89.2(3) & 139.9(4) \\
200 & 1.896(3)   & 90.6(3)/89.4(3) & 140.2(4) \\
100 & 1.901(2)   & 91.2(2)/88.8(2) & 139.4(3) \\
\hline
\end{tabular}
\end{center}
\end{table}

Figure~\ref{fig:2}b depicts another view of the crystal structure, showing that the K atom is located at the center of a cavity formed by an Os$_{12}$O$_{18}$ cage with six nearest and 12 next-nearest oxygen neighbors. The cavity is open along four directions parallel to the $\langle 111 \rangle$ direction. Thus, it can be assumed that four `hypothetical' K--K bonds pass through these channels. The sublattice of K atoms is the ideal diamond lattice, with a long K--K distance of 4.4~\AA. The diameter of the bottleneck of the channel formed by six oxygen atoms is approximately 2.4~\AA\ and is smaller than the size of the K ion (approx. 3.0~\AA).

\begin{figure}[tbh]
\centering
\includegraphics[height=5.0cm,width=7.8cm]{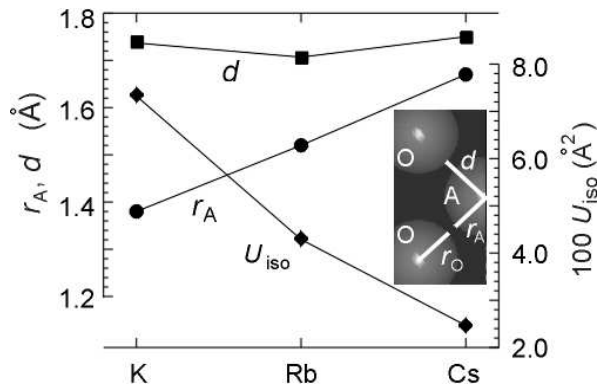}
     \caption{Ionic radius $r_\mathrm{A}$~(\AA) ($\bullet$), the cavity radius of the cage $d$~(\AA) ($\blacksquare$), and the isotropic atomic displacement parameter $U_\mathrm{iso}$~(\AA$^2$) ($\blacklozenge$) for A = Cs, Rb and K in AOs$_2$O$_6$~\cite{Yonezawa4,Shannon}.}
     \label{fig:3}
\end{figure}

The most important finding in the present structural analysis is an unusually large atomic displacement parameter (ADP) $U_\mathrm{iso}$ for the K atom, at approximately 17- (14-) and six- (four-)times greater than that for the Os and O atoms at 300~K (100~K), respectively. In previous structural analyses of CsOs$_2$O$_6$ and RbOs$_2$O$_6$, it was found that the A cations also exhibit large ADPs at room temperature, with $U_\mathrm{iso}$ = 0.025~\AA$^2$ for Cs and $U_\mathrm{iso}$ = 0.043~\AA$^2$ for Rb~\cite{Yonezawa4}. However, the present value of $U_\mathrm{iso}$ = 0.074~\AA$^2$ for K is much greater. It is apparent from Fig.~\ref{fig:3} that the smaller the ionic radius of the A cations, the larger is $U_\mathrm{iso}$ in AOs$_2$O$_6$. Similarly, large $U_\mathrm{iso}$ values have been observed in defect pyrochlore oxides (e.g. $U_\mathrm{iso}$ = 0.070~\AA$^2$ for Rb in RbNbWO$_6$)~\cite{Barnes}, filled skutterudites (e.g. $U_\mathrm{iso}$ = 0.051~\AA$^2$ for Nd in NdOs$_4$Sb$_{12}$)~\cite{Evers} and clathrates (e.g. $U_\mathrm{iso}$ = 0.033~\AA$^2$ for Ba in Ba$_8$Ga$_{16}$Si$_{30}$)~\cite{Qiu}. In all the above compounds, the large ADPs are commonly interpreted as resulting from rattling of an atom in an oversized cage. It should be noted that the K atom in KOs$_2$O$_6$ has the largest ADP value.

To gain an intuitive insight into the systematic changes in ADP observed for Cs, Rb and K in AOs$_2$O$_6$, we estimated the `size' of the free space inside the Os$_{12}$O$_{18}$ cage. A simple measure is the clearance $d$$-$$r_\mathrm{A}$, where $r_\mathrm{A}$ is the ionic radius of A and $d$ is a distance obtained by extracting the ionic radius of oxygen (1.4~\AA) from the experimental A--O distance, as schematically shown in Fig.~\ref{fig:3}~\cite{Shannon}. It was found that $d$ is almost insensitive to changes in A, implying that the Os--O framework is rigid. Thus, the clearance increases with decreasing $r_\mathrm{A}$ from Cs to K. It is obvious that $U_\mathrm{iso}$ increases with increasing clearance, which determines the space available  for the A atom. Therefore, the K atom is the most weakly bound to the surrounding atoms and can rattle heavily.

Such a weakly bound atom in an oversized cage should induce dynamic instability. Kune\v{s} et~al. found an anharmonic potential for the A cation along the $\langle 111 \rangle$ direction in AOs$_2$O$_6$ and larger anharmonicity for a smaller A cation~\cite{Kunes}. In fact, the electron distribution of the K atom shown in the Fourier map of Fig.~\ref{fig:1} is considerably elongated along the $\langle 111 \rangle$ direction, suggesting that there are tetrahedrally shaped lobes pointing to the four open channels of the Os$_{12}$O$_{18}$ cage. It is difficult to explain this elongation at the 8$b$ site within the harmonic treatment of thermal vibrations because of the high symmetry of the site in the expression of the anisotropic displacement parameters ($U_\mathrm{11}$ = $U_\mathrm{22}$ = $U_\mathrm{33}$ and $U_\mathrm{12}$ = $U_\mathrm{13}$ = $U_\mathrm{23}$ = 0). This is clear evidence for the existence of an anharmonic potential for the K atom at 300~K, which remains even at 100~K.

\begin{figure}[htb]
\centering
\includegraphics[height=5.1cm,width=7.4cm]{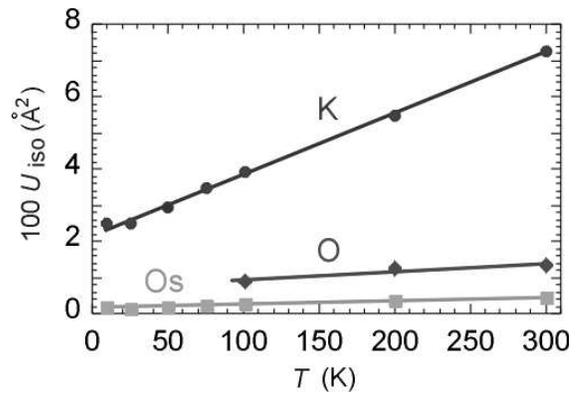}
     \caption{Temperature dependence of the isotropic atomic displacement parameter $U_\mathrm{iso}$ for K ($\bullet$), Os ($\blacksquare$) and O ($\blacklozenge$).}
     \label{fig:4}
\end{figure}

The large ADP observed can be ascribed to either dynamic atomic vibrations or static disorder. However, the previous experimental evidence that there is a large contribution from Einstein modes to the specific heat may exclude the latter possibility~\cite{Hiroi6}. It is reasonable to treat the rattling species as an Einstein oscillator and the framework as a Debye solid. Provided that the two contributions are independent of each other, characteristic temperatures $\Theta_\mathrm{E}$ and $\Theta_\mathrm{D}$ can be estimated for the two modes from the temperature dependence of the mean-square displacement amplitude $\langle u^2 \rangle = U_\mathrm{iso}$. Figure~\ref{fig:4} shows the temperature dependence of $U_\mathrm{iso}$ for K, Os and O atoms. The $U_\mathrm{iso}$ values for K and Os below 100~K were calculated from the intensity data of some selected reflections measured by the imaging plate. The mean-square displacement amplitude is given by $\langle u^2 \rangle = u_0^2+h^2/(8\pi^2mk_\mathrm{B}\Theta_\mathrm{E})\mathrm{coth}(\Theta_\mathrm{E}/2T)$ for a harmonic oscillator, and $\langle u^2 \rangle = 3h^2/(4\pi^2mk_\mathrm{B}\Theta_\mathrm{D})T$ ($T>\Theta_\mathrm{D}$) for a Debye solid, where $u_0$ is a temperature-independent parameter, $h$ is Planck's constant, $k_\mathrm{B}$ is the Boltzmann constant, and $m$ is the atomic mass~\cite{Sales1,Willis}. The Einstein temperature $\Theta_\mathrm{E}$ is estimated to be 86(2)~K by fitting the data for K over the whole temperature range, and the Debye temperature $\Theta_\mathrm{D}$ is estimated to be 266(8)~K using the average $U_\mathrm{iso}$ value and the average mass per atom for Os and O at 300~K. This $\Theta_\mathrm{E}$ value is close to the average value of two Einstein temperatures obtained from specific heat measurements~\cite{Hiroi6}. The temperature-independent parameter was found to be very large: $u_0$ = 0.147(5)~\AA. Generally, it is ascribed to certain disorder, but this may not be the case now. The large $u_0$ value may indicate that the rattling of the K atom remains dynamic, even at low temperature, owing to the anharmonic potential with an almost flat bottom for the K atom~\cite{Kunes}.

Previous specific heat measurements detected an intensive peak centered at 7.5~K below $T_\mathrm{c}$, which was not directly ascribed to superconductivity, but might be related to an unknown structural transition~\cite{Hiroi7}. We carried out a preliminary structure analysis at 5~K. No evidence of a structural phase transition was detected. However, a further study is required to clarify this interesting possibility.

\section{Summary}
In summary, the structure of KOs$_2$O$_6$ has been determined as the $\beta$-pyrochlore structure. The K atom is located at the 8$b$ site, not at the 16$d$ site as in conventional pyrochlore oxides, and thus at the center of the Os$_{12}$O$_{18}$ cage. An anomalously large atomic displacement parameter was found for the K atom, which is because of rattling of the K atom in the oversized cage. It has been shown that the magnitude of rattling in $\beta$-pyrochlore oxides is determined by the clearance between the cavity of the cage and the cation: the greater the clearance from Cs to K, the greater is the rattling magnitude. In addition, it was found that the K atom vibrates with anharmonicity along the $\langle 111 \rangle$ direction to the open channels of the Os$_{12}$O$_{18}$ cage.

\end{document}